\documentclass[twocolumn,prl,showpacs,floatfix]{revtex4}

\usepackage{subfigure}
\usepackage{graphicx}

\begin{document}

\preprint{DRAFT}
\date{July 31, 2018}

\title{Axion dark matter and the 21-cm signal}
\author{Pierre Sikivie}

\affiliation{Department of Physics, University of Florida, 
Gainesville, FL 32611, USA}

\begin{abstract}

It was shown in ref. \cite{Erken} that cold dark matter axions 
reach thermal contact with baryons, and therefore cool them, 
shortly after the axions thermalize among themselves and 
form a Bose-Einstein condensate. The recent observation by the 
EDGES collaboration of a baryon temperature at cosmic dawn lower 
than expected under ``standard" assumptions is interpreted as 
new evidence that the dark matter is axions, at least in part.  
Baryon cooling by dark matter axions is found to be consistent 
with the observation of baryon acoustic oscillations.

\end{abstract}

\maketitle

The EDGES collaboration reported recently \cite{EDGES} the 
observation of the trough in the spectrum of cosmic microwave
radiation caused by its absorption by neutral hydrogen atoms 
\cite{Furlanetto} at cosmic dawn, i.e. when the universe is 
bathed in starlight for the first time.  Assuming it is correct, 
the EDGES observation reveals new important information.  
It informs us that the first stars formed approximately 
180 million years after the Big Bang and that the primeval 
gas was heated to above the photon temperature approximately 
100 million years later.  Most importantly for the present 
discussion, it tells us the spin temperature of hydrogen 
atoms during this 100 million year epoch.  The spin 
temperature $T_s$ is, by definition, related to the relative 
population of the spin 1 and spin 0 lowest energy states of 
hydrogen: ${n_1 \over n_0} \equiv 3 \times e^{- \omega_0/T_s}$ 
where $\omega_0 = (2 \pi)~$ 1.42 GHz is the angular frequency 
associated with hyperfine splitting in hydrogen, the so-called 
21 cm line.  We use units in which $\hbar = c = k_{\rm B} = 1$ 
throughout.  The spin temperature is intermediate between the 
baryon kinetic temperature $T_k$ and the photon temperature 
$T_\gamma$ ($T_k \leq T_s \leq T_\gamma$) being driven toward 
$T_\gamma$ by 21 cm emission and absorption but driven towards 
$T_k$ by atomic collisions and by Lyman-alpha emission and absorption. 
The latter process is known as the Wouthuysen-Field effect \cite{WF}. 
After recombinaton, $T_k$ and $T_\gamma$ continue to be kept equal 
by the action of a residual population of free electrons that 
Thompson scatter with photons and scatter with atoms through 
Coulomb forces.  However after redshift $z_{\rm dec} \sim 160$, 
these processes become ineffective, baryons and photons decouple, 
and the baryons cool relative to the photons since 
$T_k \propto a(t)^{-2}$ for decoupled baryons whereas 
$T_\gamma \propto a(t)^{-1}$ for decoupled photons. At 
$z_{\rm cd} \simeq$ 17.2, which corresponds to the middle of 
the absorption trough observed by EDGES, the temperature of 
baryons expected under ``standard" assumptions is near 7.6 K.  
On the other hand, the EDGES measurement indicates 
1.73 K $< T_s <$ 5.34 K at 99\% confidence level, and 
hence $T_k \leq 5.34$ K. A perhaps related puzzle concerns 
the profile of the absorption trough:  the  profile observed 
by EDGES is more flattened than had been predicted \cite{Cohen}.

It is appropriate to ask what is the degree of discrepancy between 
the expected $T_k \simeq 7.6$ K and the measured $T_s < 5.34$ K, 
granted that $T_k \leq T_s$.  The predicted value depends on the
density of free electrons after recombination.  It is thought that 
this parameter is known with a precision of order one percent as 
a result of cross-validation between numerical models and Planck 
observations \cite{EDGES}.  If so, this uncertainty is too small 
to explain the discrepancy.  Ref. \cite{Fialkov} presents the 
range of predictions for the 21 cm observations consistent 
with the plausible values of the relevant cosmological 
parameters.  The measured values of the redshifts at which 
the absorption trough begins and ends are consistent with 
the theoretical expectations. The depth of the trough, which 
is inversely proportional to $T_s$ and proportional to the 
amount of radiation at cosmic dawn with frequency near 1.42 GHz, 
is approximately double the expected value.  The discrepancy 
is estimated to be 3.8 $\sigma$ \cite{Barkana}.
 
One may attempt to explain the discrepancy by postulating 
additional radiation at the time of cosmic dawn with frequency 
near 1.42 GHz.  The discrepancy is removed if the total amount 
of radiation is double the amount expected from the cosmic 
microwave background.  The additional radiation may be 
synchrotron emission associated with the first stars, 
supernovae and black holes but the efficiency of these 
sources as radio emitters would have to be three orders 
of magnitude larger than their low redshift counterparts.  
An additional problem is that the cooling time of the 
synchrotron emitting electrons is 3 orders of magnitude 
shorter than the duration of the absorption signal \cite{Sharma}.
 
The most widely discussed explanation of the discrepancy is 
that baryons are cooled by thermal contact with cold dark matter 
\cite{Barkana,Fialkov,Barkana2}. The dark matter candidates that 
have been considered in this context are weakly interacting 
massive particles (WIMPs) with however much lower mass than 
WIMPs had heretofore been thought to have and with Coulomb-type 
interactions (scattering cross-section $\sigma \propto v^{-4}$ 
where $v$ is the relative velocity between the scatterers) 
with ordinary matter.  Such interactions are present if the 
WIMP has a small electric charge or if it couples to a dark 
photon which mixes with the ordinary photon.  Dark matter with 
such properties is commonly referred to as "milli-charged" dark 
matter.  It is constrained by a large number of considerations,
including its absence from a dedicated search at SLAC and its
effect on Supernova 1987a, on the cosmic microwave background 
observations, and on primordial nucleosynthesis.  When the 
various constraints are taken into account \cite{Munoz,Hooper}, 
the only range of models that may explain the EDGES 21-cm signal 
are those in which a small fraction, from 0.3 to 2 \%, of the 
dark matter is particles with mass from 1 to 100 MeV and 
which couple to the photon with an electric charge of order 
$10^{-4}$ to $10^{-6}$ that of the electron. Such particles 
would tend to be overproduced in the early universe and 
would need to be depleted in some way.

The purpose of the present paper is to point out that if 
the dark matter is axions, baryons are cooled automatically 
in the cosmological context, without the need of additional 
assumptions.  Baryon cooling by Bose-Einstein condensed axions 
\cite{CABEC} was predicted \cite{Erken} and is generic \cite{Banik} 
to all (pseudo)-scalar dark matter produced in the early universe 
through the vacuum-realignment mechanism \cite{axdm}, including 
QCD axions \cite{PQWW}, axion-like particles \cite{ALPs} and 
ultralight axion-like particles \cite{ULALP}.  The relevant 
interaction is gravity.

Cold dark matter axions thermalize in a regime where their 
energy dispersion $\delta \omega = {1 \over 2m}(\delta p)^2$
is much less than their thermalization rate $\Gamma$.   In
this regime, called the ``condensed regime", the axion
thermalization rate by gravitational self-interactions is
\begin{equation}
\Gamma \sim 4 \pi G n m^2 \ell^2 
\label{rr2}
\end{equation}
where $\ell = 1/\delta p$ is the correlation length.  
Eq.~(\ref{rr2}) was derived analytically in ref.\cite{Erken}, 
and the derivation was validated by numerical simulation of a 
toy model \cite{Erken,simtherm}.  At the QCD phase transition 
when the axion mass turns on, $\Gamma$ is much less than $H$.  
However, $\Gamma/H$ grows with time since $\Gamma \propto a(t)^{-1}$ 
whereas $H(t) \propto a(t)^{-2}$.  For the case where inflation 
occurs before the Peccei-Quinn phase transition and the axion mass 
is near $10^{-5}$ eV, $\Gamma \sim H$ when the photon temperature 
$T_\gamma$ is near 500 eV. The axions thermalize then. In all 
cases, axion or ALP dark matter produced by the vacuum realignment 
mechanism thermalizes at or before approximately the time of 
equality between matter and radiation \cite{Banik}.

When the axions thermalize, all conditions for their Bose-Einstein
condensation are satisified and we should expect that this is indeed 
what happens.  Bose-Einstein condensation means in this highly 
degenerate case that almost all axions go to the lowest energy 
state available to them by the thermalizing interactions.  The 
remaining axions form a thermal or quasi-thermal distribution, whose 
temperature $T_a$ can be obtained by energy conservation.  In the 
aforementioned case where the axion mass is $10^{-5}$ eV, $T_a$ 
is of order $10^{-4}$ eV at the time of equality between matter 
and radiation assuming the axions thermalize completely. The 
complete thermalization of axions generally takes much more 
time than $\Gamma^{-1}$.  Axion Bose-Einstein condensation 
however takes place on the $\Gamma^{-1}$ time scale.  The 
correlation length $\ell$ of the Bose-Einstein condensed 
axions grows without bound except that its ultimate value 
must be less than $t$ by causality.  

Axion Bose-Einstein condensation is a difficult topic from 
a theoretical point of view.  The idea that dark matter 
axions form a Bose-Einstein condensate was critiqued in refs. 
\cite{Davidson,Davidson2,Guth}.  The main difficulty is that 
the axions thermalize by gravitational self-interactions 
and gravity causes instability since it is attractive.  
One point of contention is whether a Bose-Einstein 
condensate must be homogeneous.  Bose-Einstein condensation 
means that a macroscopically large number of particles go 
to the lowest energy state available to them through the 
thermalizing interactions. That state need not be, and in 
general is not, homogeneous.  In the axion case, it is not
only inhomogeneous in general but also time-dependent. 
Whether homogeneous or not, a Bose-Einstein condensate is 
correlated over its full spatial extent \cite{SC}. Another 
point of contention is whether Bose-Einstein condensation 
can be described by classical field theory.  When a cutoff 
on the wavevectors is introduced, a phenomenon akin to 
Bose-Einstein condensation may occur in classical field 
theory.  However this does not mean that the cutoff classical 
field theory has validity beyond the built-in property of 
producing some form of Bose-Einstein condensation.  The
axion field has no wavevector cutoff. The classical field 
theory, with or without cutoff, differs from the full quantum 
field theory in important ways.  In particular, it conserves 
vorticity whereas the quantum field theory does not \cite{angmom}.  
Refs. \cite{simtherm,SC} gives a detailed discussion of the 
issues related to cosmic axion Bose-Einstein condensation.

It was shown in ref. \cite{Erken} that gravitational 
scattering of axions with baryons produces a deceleration
\begin{equation}
d \sim 4 \pi G n m \ell~~\ .
\label{dec}
\end{equation}
of any baryon moving with respect to the axion fluid.
The deceleration is velocity-independent.  It depends 
only on the product of the density and the correlation 
length of the axion fluid.  Using the Friedmann equation
we may write 
\begin{equation}
d \sim H(t)~Y(t)~{\ell \over t}
\label{dec2}
\end{equation}
where $H(t)$ is the Hubble rate and $Y(t) = m n(t)/\rho(t)$ 
the ratio of the energy density in axions to the total 
energy density.  

The axions may have a cooling effect on photons \cite{Tam,Erken}.
The relative rate $- \dot{\omega}/\omega$ at which a photon decreases 
its energy $\omega$ is also of order the RHS of Eq.~(\ref{dec2})
\cite{Erken}. Photon cooling by Bose-Einstein condensed was considered 
\cite{Tam} as a possible explanation of the Lithium anomaly in 
primoridial nucleosynthesis.  It would occur around the time of 
equality if $Y(t)~{\ell \over t}$ is of order one, its largest 
possible value.  If it did occur, there would be a substantial
increase in the effective number $N_{\rm eff}$ of relativistic number 
of degrees of freedom.  Since a large increase in $N_{\rm eff}$ is 
inconsistent with the Planck observations, not much photon cooling 
occurred implying that $Y(t) {\ell(t)/t}$ is much less than one, 
perhaps 0.01 or less.  A precise limit has not been established.

Axion Bose-Einstein condensation explains \cite{case}, in detail 
and in all aspects, the observational evidence for caustic 
rings of dark matter \cite{crdm,singul} in galactic halos.  The
relevance of dark matter caustics to astronomical observation and
direct detection on Earth was questioned in refs. \cite{Nbody} 
on the basis of N-body numerical simulations.  However, the 
resolution of present day N-body simulations is inadequate
to resolve the detailed phase structure of galactic halos.  
Because a typical simulated particle weighs $10^5~M_\odot$, 
the number of particles per halo is of order $10^8$ or less.   
In contrast, the typical mass of cold dark matter candidates 
is $10^{-55}~M_\odot$ in the case of weakly interacting massive 
particles (WIMPs), and $10^{-71}~M_\odot$ in the case of axions.  
Because of their huge numbers, WIMPs and axions lie on a continuous 
3-dimensional hypersurface in phase-space.  This hypersurface 
is poorly sampled in the simulations.  Indeed, since phase-space 
is six-dimensional, there are only $10^{8/6} \sim$ 22 particles 
per phase space dimension in the simulated halos, making it 
impossible to see but a very few of the expected caustics.  
Arguments for the robustness of discrete flows and caustics 
in cold dark matter halos were given in ref. \cite{robust}.
Note at any rate that, even if the present simulations had 
high enough resolution, they would not produce caustic rings 
because caustic rings require the production of vorticity.
The production of vorticity is a quantum effect associated
with Bose-Einstein condensation \cite{angmom}.  That physics
is not included in the simulations.

The observational evidence for caustic ring of dark matter is 
summarized in ref. \cite{Duffy}; see also ref. \cite{Dumas}.  All 
aspects of that phenomenolgy is explained by axion Bose-Einstein 
condensation and rethermalization, including the catastrophe 
structure of the caustic rings, the fact that they lie in the 
galactic plane, the $a_n \propto {1 \over n}~ (n = 1, 2, ,3 ..)$ 
pattern of the caustic ring radii, and the overall size of 
the rings \cite{case}. The explanation requires that axions 
rethermalize sufficiently fast so as to acquire net overall 
rotation before falling onto galactic halos.  For this to 
occur the axions must be at least of order 3 \% of the dark 
matter, $Y \gtrsim 0.01$, and their correlation length at 
least galaxy size, i.e. $\ell/t \gtrsim 10^{-3}$.  The 
entrainment of baryons by the rotating axions was shown to 
solve \cite{angmom} the galactic angular momentum problem.  
Ref.~\cite{angmom} provides an argument, based on the prominence 
of caustic rings, that the axion fraction of dark matter is at 
least of order 0.375.  In summary the evidence for caustic rings 
requires that $Y \ell/t$ is larger than $10^{-5}$ or $10^{-4}$.

Baryon cooling by axion BEC is very efficient in the absence
of a competing process heating the baryons up.  Consider 
baryons at temperature $T_k$.  Their typical velocities 
are $v \sim \sqrt{3T_k/m_B} \sim$ 500 m/s $\sqrt{T_k/10~{\rm K}}$.  
On the other hand, if $Y \ell/t \sim 10^{-5}$, their deceleration 
is of order $10^{-5} H \sim 3 \cdot 10^3$ m/s per Hubble time.
This implies that shortly before cosmic dawn, the baryon 
temperature is essentially zero, much less at any rate than 
under the standard assumptions.  At cosmic dawn, the baryons 
are heated up by the radiation from the first stars.  Perhaps 
the spin temperature at the bottom of the absorption trough
equals the baryon temperature then.  We note however that 
the absorption trough observed by EDGES appears saturated.  
The optical depth for absorption of 21 cm radiation by 
neutral hydrogen is \cite{Furlanetto}
\begin{equation}
\tau \simeq 0.0092~x_{H1}~(1 + z)^{3 \over 2} {{\rm K} \over T_s}
\label{od}
\end{equation}
where $x_{HI}$ is the neutral hydrogen fraction.
For $z = z_{cd} = 17.2$, $x_{HI} = 1$, and $T_s = 3.35$ K,
the spin temperature implied by the depth of the EDGES 
absorption trough, $\tau \simeq 0.2$.  This fairly large 
value  suggests that absorption of 21 cm radiation, which 
drives $T_s$ up, may compete with the Wouthuysen-Field 
effect driving it down.  In that case the absorption 
would saturate before $T_s$ reaches $T_k$. 

Whether baryon cooling by dark matter axions agrees 
with all observations remains to be seen.  Eq.~(\ref{dec})
allows predictions to be made.  One concern is whether 
the drag on baryons dampens baryon acoustic oscillations.
The damping of baryon acoustic oscillations by WIMP 
dark matter was considered in refs. \cite{BAO} and 
constraints on the scattering cross-section between 
WIMPs and baryons were derived on this basis.   
Eq.~(\ref{dec2}) implies that the fraction of energy 
removed from the baryon-photon fluid per Hubble time 
due to the drag of baryons on the axion fluid is of 
order 
\begin{equation}
{1 \over \rho_\gamma~H}{d \rho_B \over dt} \simeq
{m_B v d \over (2.7) T_{\rm rec} H}
{n_B \over n_\gamma}
\sim 3 \cdot 10^{-5} Y {\ell \over t}
\label{bao}
\end{equation}
shortly before recombination.  We used $m_B =$ GeV 
for the baryon mass, $T_{\rm rec} = 0.256$ eV for 
the temperature at recombination, 
$v = \sqrt{3 T_{\rm rec} \over m_B}$ for the typical 
baryon speed at that time (the bulk velocities associated 
with density perturbations are subdominant), and 
${n_B \over n_\gamma} = 6.1 \cdot 10^{-10}$ for 
the baryon to photon ratio.  In the expected range 
$10^{-2} < Y {\ell \over t} < 10^{-5}$, the drain 
of energy indicated by Eq.~(\ref{bao}) is far too 
small to affect the observations of baryon acoustic 
oscillations.  

While the original version of the present paper 
\cite{Tmat1} was being written up, ref. \cite{Houston} 
on the same topic appeared.  Ref.~\cite{Houston} also 
interprets the EDGES observations as the result of the 
cooling of baryons by Bose-Einstein condensed axions but 
makes additional assumptions.  In particular the assumption 
that cooling by axion BEC does not cause the baryon kinetic 
temperature to deviate from the photon temperature before 
$z_{\rm dec} \sim 160$ leads the authors to predict a large 
value for the axion mass, of order 0.15 eV \cite{Houston}.  

I thank Aravind Natarajan, Vera Gluscevic and Wei Xue 
for making me aware of various aspects of 21 cm physics.
I gratefully acknowledge the hospitality and support of the  
Kavli Institute for Theoretical Physics at UC Santa Barbara
while working on this paper, and the stimulationg conversations 
I had there with Graciela Gelmini, Jordan Mirocha and Steven 
Furlanetto.  This research was supported in part by the U.S. 
Department of Energy under grant DE-FG02-97ER41029, by the 
National Science Foundation under grant NSF PHY11-25915, and 
by the Heising-Simons Foundation under grant No. 2015-109.

\end{document}